\documentstyle[11pt]{article}

[1999/03/29 PSNFSS v.7.2
Times font as default roman
: S Rahtz]

\newcommand \be{\begin{equation}}
\newcommand \ba{\begin{eqnarray}}
\newcommand \ee{\end{equation}}
\newcommand \ea{\end{eqnarray}}

\author{S. Gluzman$^{1}$ and D. Sornette$^{1,2,3}$\\
$^1$ Institute of Geophysics and Planetary Physics\\
University of 
California Los Angeles, Los Angeles, CA 90095-1567\\
$^2$ Department of Earth and Space Sciences, UCLA\\
$^3$ Laboratoire de Physique de la Mati\`ere Condens\'ee\\
CNRS UMR 6622 and Universit\'e de Nice-Sophia Antipolis\\
06108 Nice Cedex 2, France}
\title{\LARGE Classification of Possible Finite-Time Singularities by  Functional Renormalization
}
\begin{document}

\maketitle
\begin{abstract}
Starting from a representation of the early time evolution of a dynamical
system in terms of the polynomial expression of some observable $\phi (t)$
as a function of the time variable in some interval $0\leq t\leq T$, we
investigate how to extrapolate/forecast in some optimal stability sense the
future evolution of $\phi (t)$ for time $t>T$. Using the functional
renormalization of Yukalov and Gluzman, we offer a general classification of
the possible regimes that can be defined based on the sole knowledge of the
coefficients of a second-order polynomial representation of the dynamics. In
particular, we investigate the conditions for the occurence of finite-time
singularities from the structure of the time series, and quantify the
critical time and the functional nature of the singularity when present. We
also describe the regimes when a smooth extremum replaces the singularity
and determine its position and amplitude. This extends previous works by (1)
quantifying the stability of the functional renormalization method more
accurately, (2) introducing new global constraints in terms of moments and
(3) going beyond the ``mean-field'' approximation.
\end{abstract}

\pagebreak

\section{Introduction}

Finite-time singularities in the dynamical equations used to describe
natural systems are not always pathologies that should be thrown away or
ignored but may often betray important useful information on the
characteristic properties of the real system. Actually, spontaneous
singularities in ordinary and partial differential equations are quite
common and have been found in many well-established models of natural
systems, either at special points in space such as in the Euler equations of
inviscid fluids \cite{Pumir,eulerstag}, in the surface curvature on the free
surface of a conducting fluid in an electric field \cite{Zubarev}, in vortex
collapse of systems of point vortices \cite{Zaslav}, in the equations of
General Relativity coupled to a mass field leading to the formation of black
holes \cite{Choptuik}, in models of micro-organisms aggregating to form
fruiting bodies \cite{Rascle}, or in the more prosaic rotating coin (Euler's
disk) \cite{Moffatt,eulerdiskophys}. Some more complex examples are models
of rupture and material failure \cite{critruptcan,critrupt,herma},
earthquakes \cite{earthquake,SamsorPNAS} and stock market crashes \cite
{crash,nasdaq}.

In a recent work \cite{GASpred}, we have developed theoretical formulas for
the prediction of the singular time of systems which are a priori known to
exhibit a critical behavior, based solely on the knowledge of the early time
evolution of an observable. From the parameterization of such early time
evolution in terms of a low-order polynomial of the time variable, the
functional renormalization approach introduced by Yukalov and Gluzman allows
one to transform this polynomial into a function which is asymptotically a
power law. The value of the critical time $t_c$, {\it conditioned} on the
assumption that $t_c$ exists, can then be determined from the knowledge of
the coefficients of the polynomials. Ref.\cite{GASpred} has tested with
success this prediction scheme on a specific example and showed that this
approach gives more precise and reliable predictions than through the use of
the exact real power law model. This is a rather surprising and paradoxical
observation in contradiction with common wisdom according to which the best
and most reliable prediction should be obtained from the use of the exact
underlying model of the dynamical behavior \cite{Ziehmann}. The reason why
this is not always true is that our method has shown that approximate
solutions can be more stable and more reliable when devised so as to
maximize the criterion of stability with respect to perturbations in the
functional space \cite{GASpred}.

Here, we extend this work by offering a general classification of the
possible regimes that can be defined based on the sole knowledge of the
coefficients of the polynomial expansion of some observable as a function of
the time variable. We mostly restrict our analysis to second-order
polynomials. As a consequence, the classification can be organized in a
unique way in terms of the single signed Froude parameter, ratio of the
square of the velocity to the acceleration.

Let us assume that the dynamical behavior of a system is sampled to obtain a
time series $\phi(t)$ in the interval $0 \leq t \leq T$. The question we
address in this work is how to extrapolate/forecast in some optimal sense
the future evolution of $\phi (t)$ for time $t>T$. Beyond a simple
extrapolation, we ask whether it is possible to detect the germs of a
finite-time singularity from the structure of the time series, and quantify
the critical time and the functional nature of the singularity when present.
We also aim at classifying the regimes when a smooth extremum replaces the
singularity and at determining its position and amplitude.

We shall work in the framework of the functional renormalization method,
which constructs the extrapolation for $t>T$ from a resummation of the time
series represented by a simple model-free polynomial expansion in powers of
time $t$, where $t$ is counted since the beginning of the recorded time
series.

\section{Summary of the functional renormalization approach to extrapolation
of times series \label{methoconvergent}}

The mathematical foundation of the functional renormalization approach used
here can be found in earlier publications \cite{G1}-\cite{G13}, which we
summarize briefly. Let perturbation theory (or some fitting procedure) give
for the time series $\phi (t)$ the succession of non-random approximations $%
\phi _n(t)$ where $n=0,1,2,...$ enumerates the order of the increasingly
precise approximations. The case of a polynomial expansion will be discussed
below, 
\begin{equation}
\label{eq1}\phi _n(t)=\sum_{k=0}^na_k~t^k~,~~~~~~~~~n=0,1,2,3,... 
\end{equation}
This expansion (\ref{eq1}) is in principle defined for $t$ sufficiently
small and based on information about the time series up to the time $T$. The
expansion may have no meaning if continued straightforwardly to the region
of finite $t>T$. The problem of reconstructing the value of the function in
some distant moment of time from the knowledge of its asymptotic expansion
as $t\rightarrow 0$ is called a renormalization or resummation problem in
theoretical physics. An analytical tool for the solution of this problem,
called algebraic self-similar renormalization, has been developed recently 
\cite{G1}-\cite{G13} of which we summarize the salient points useful for the
present work.

It is convenient to remove the constant term and consider the time series $%
p(t)\equiv \phi (t)-\phi _0$ represented by the sequence of polynomial
approximations $p_i(t),i=1,2,...,n,$%
\begin{equation}
\label{eq2}%
p_1(t)=a_1t,~~~~~~~p_2(t)=p_1(t)+a_2t^2~,...~,~~~p_n(t)=p_{n-1}(t)+a_nt^n~. 
\end{equation}
The algebraic self-similar renormalization starts by applying to the
approximations (\ref{eq2}) a simple algebraic multiplication, thus defining
a new sequence, $P_i(t,s)=t^sp_i(t),\ i=1,2,...,n,$ with $s\geq 0$. This
transformation increases the powers of the series (\ref{eq1}) and (\ref{eq2}). 
This formal manipulation effectively increase the order of the expansion
and provides a trick to effectively take into consideration more points from
the system trajectory. In the first part of the paper, we use the strongest
form of this transformation corresponding to formally take the limit $%
s\rightarrow \infty $. This will be shown to provide a representation of the
renormalized function as an imbedded set of exponentials associated with a
singularity $\phi \sim (t_c-t)^z$, with
critical exponent $z$ imposed to have an absolute value equal to $1$. This limit
can thus be considered as analogous to a ``mean field'' regime with the mean
field value of the critical exponent $z=-1$. This can for instance represent
the exponent of the derivative of a quantity exhibiting a weak logarithmic
divergence. In the second part of the paper, we shall relax this limit and
determine both $t_c$ and the index $z$ self-consistently.

The second step consists in considering the sequence of transformed
approximations $P_i(t,s),$ as a dynamical system in the discrete ``order
time'' equal to the order $i=0,1,...,n-1$ of the approximation. In order to
keep the information on the system evolution with the real time $t$, it is
convenient to introduce a new variable $\varphi $ and define the so-called
expansion function $t(\varphi ,s)$ from the equation $P_1(t,s)=a_1t^{1+s}=%
\varphi $, which gives $t(\varphi ,s)=\left( \varphi /a_1\right) ^{1/1+s}$.
We then construct the discrete flow in the space of approximations indexed
by the ``order time'' as 
\be
y_i(\varphi ,s)\equiv P_i(t(\varphi ,s),s)~. 
\ee
One can then write the equation of evolution in the space of approximations
as a function of the discrete ``order time'' in the form of the functional
self-similarity relation \be
y_{i+p}(\varphi ,s)=y_i(y_p(\varphi ,s),s)~. \label{eq3} \ee
Expression (\ref{eq3}) provides the necessary condition for the
self-consistency of the cascade of discrete approximations and ensures the
convergence of $P_i$'s. Expression (\ref{eq3}) is nothing but the
application of the Caccioppoli-Banach principle for the existence of a
stable fixed point (see for instance Chapter XVI of \cite{kantoro}).

At this stage, the efficiency of the algebraic transformation can be checked
by analyzing its stability. This is done for the sequence of $P_i(t,s)$'s by
calculating the so-called local multipliers (essentially proportional to the
exponential of Lyapunov exponents) 
\begin{equation}
\label{eq4}m_i(t,s)\equiv [\frac{\partial y_i(\varphi .s)}{\partial \varphi }%
]_{\varphi =P_1(t,s)}~. 
\end{equation}
When all $\left| m_i(t,s)\right| <1,$ the convergence of the sequence $P_i$
is guaranteed. To implement concretely the calculations, we use the integral
form of the self-similarity relation (\ref{eq3}) \be
\int_{P_{i-1}}^{P_i^{*}}\frac{d\varphi }{v_i(\varphi ,s)}=\tau ~, \ee
where the cascade velocity is $v_i(\varphi ,s)=y_i(\varphi
,s)-y_{i-1}(\varphi ,s)$ and $\tau $ is the minimal number of steps of the
approximation procedure needed to reach the fixed point $P_i^{*}(t,s)$ of
the approximation cascade. It is possible to find $P_i^{*}(t,s)$ explicitly
and to perform an inverse algebraic transform after which the limit $%
s\rightarrow \infty $ is to be taken. This completes a first loop of the
self-similar renormalization. This procedure can be repeated as many times
as it is necessary to renormalize all polynomials which appear at the
preceding steps. This is the main idea of the self-similar bootstrap \cite
{G5}.

Completing this program, we come to the following sequence of self-similar
exponential approximants 
\begin{equation}
\label{eq5}p_j^{*}(t,\tau _{1,}\tau _{2,}...,\tau _{\ j-1})=a_1t\exp \left( 
\frac{a_2}{a_1}t\tau _1...\exp \left( \frac{a_j}{a_{j-1}}\tau _{\
j-1}t\right) \right) ...,\ \ j=2,3...,n~. 
\end{equation}
and \be
\phi _j^{*}(t,\tau _{1,}\tau _{2,}...,\tau _{\ j-1})=p_j^{*}(t,\tau
_{1,}\tau _{2,}...,\tau _{\ j-1})+a_0~. \ee

Explicitly, for the three first orders, we obtain \be
\phi _2^{*}(t,\tau _1)=a_1t\exp \left( \frac{a_2}{a_1}t\tau _1\right) +a_0~, 
\ee
\be
\phi _3^{*}(t,\tau _1,\tau _2)=a_1t\exp \left( \frac{a_2}{a_1}t\tau _1\exp
\left( \frac{a_3}{a_2}t\tau _2\right) \right) +a_0~, \ee
\be
\phi _4^{*}(t,\tau _1,\tau _2,\tau _3)=a_1t\exp \left( \frac{a_2}{a_1}t\tau
_1\exp \left( \frac{a_3}{a_2}t\tau _2\exp \left( \frac{a_4}{a_3}t\tau
_3\right) \right) \right) +a_0~. \ee

Compared to previous works \cite{G1}-\cite{G13}, we introduce the following
innovation to improve on the selection of the stable extrapolations (or
scenarios). In order to check whether the sequence\ of $\phi _j^{*}(t,\tau
_{1,}\tau _{2,}...,\tau _{\ j-1})$ really converges, in addition to
calculating the multipliers $m_i(t,s)$ defined by (\ref{eq4}), we analyze
the multipliers $M_j(t,\tau _{1,}\tau _{2,.._{\ },}\tau _{\ j-1})$
corresponding specifically to $\phi _j^{*}(t,\tau _{1,}\tau _{2,}...,\tau
_{\ j-1})$. For this, we construct again an approximation cascade as
described above and define 
\begin{equation}
\label{eq6}M_j(t,\tau _{1,}\tau _{2,.._{\ },}\tau _{\ j-1})\equiv [\frac{%
\partial p_j^{*}(\varphi ,\tau _{1,}\tau _{2,.._{\ },}\tau _{\ j-1})}{%
\partial \varphi }]_{_{\varphi =P_1(t,0)}}~. 
\end{equation}
Such a definition of multipliers allows us for the first time to compare the
convergence of the expansion and of the renormalized expansion, making clear
the improvements that can be expected a priori from the technique. If the
values of $M_j$ are systematically smaller than the corresponding $m_j$ in
the region of $t\leq T$ and the overall convergence properties are improved,
one can expect that the renormalized expressions will work better than the
original expansion at $t>T$.

The final step consists in determining the control parameters $\tau
_{1,}\tau _{2,\ ..._{\ },}\tau _{\ j-1}$, by expanding $\phi _j^{*}(t,\tau
_{1,}\tau _{2,}...,\tau _{\ j-1})$ in the vicinity of $t=0$, and requiring
that this expansion agrees term-by-term with the initial one in terms of the 
$\phi _j(t)$. For each $j$, we obtain $j-1$ self-similar approximants for
the sought function (where all control parameters $\tau $ are now known
functions of the parameters $a$, with $\tau _1\equiv 1$), \ba
\phi _{j1}^{*}(t,1,1...,1)~, \nonumber \\ \phi _{j2}^{*}(t,1,\tau
_{2,}1,...,1)~, \nonumber \\ ... \\ \phi_{j~j-1}^{*}(t,1,\tau _{2,}...,\tau
_{j-1})~, \label{mgnglqw} \ea
which differ according to the number of control functions being used. We can
now construct a table of self-similar approximants, varying both $j$ and the
number of controls. Accordingly, we can define the table of multipliers,
varying both $j$ and the number of controls. For instance, for $j=4,$ we
have the following table of approximations: 
$$
\phi _{21}^{*}(t)=\phi _j^{*}(t,1)~,\qquad \qquad \quad \quad \quad \quad
\qquad \qquad \qquad \qquad \qquad \ \qquad 
$$
$$
\phi _{31}^{*}(t)=\phi _3^{*}(t,1,1),\ \phi _{32}^{*}(t)=\phi
_3^{*}(t,1,\tau _{2,})~,\qquad \quad \quad \quad \quad \qquad \qquad 
$$
$$
\qquad \phi _{41}^{*}(t)=\phi _4^{*}(t,1,1,1),\ \phi _{42}^{*}(t)=\phi
_4^{*}(t,1,\tau _2,1),\ \phi _{43}^{*}(t)=\phi _4^{*}(t,1,\tau _2,\tau _3)~. 
$$

In order to select the scenarios for the extrapolation to the future from
the initial time series, one needs to examine both the properties of
convergence of the sequences of multipliers $M$, and of approximants $\phi
^{*}$. The best scenario is the one which exhibits the best simultaneous
convergence of both sequences. The resulting limiting fixed point read from
the table of approximants should then be taken for the sought extrapolation
function. If there are more than one limiting points, the sought function
should be constructed by taking their average with weights inversely
proportional to the multipliers (see \cite{G13}). We now illustrate this
general methodology on specific examples for which a finite-time singularity
might exist.

\section{Classification of singular and non-singular behaviors based on
second order expansions \label{seckgja}}

\subsection{Definitions}

Let us now consider the simplest non-trivial case allowing for the possible
existence of a finite-time singularity, namely a second-order polynomial
representation 
\begin{equation}
\label{eq7}\phi_2(t)\simeq 1+a_1t+a_2t^2\ \quad (t\rightarrow 0) 
\end{equation}
of the initial time series. In the language of \cite{Techanal}, $a_1$ is the
``velocity'' and $2a_2$ is the ``acceleration''. The relative influence of
the velocity and acceleration is quantified by the so-called Froude number 
\cite{Techanal} defined by 
\begin{equation}
\label{eq8}F_d\equiv {\frac{a_1^2}{a_2}}~. 
\end{equation}
Working with the ``direct'' series (\ref{eq7}), the program described in the
previous section provides the sought observable, that we name $\Phi_2^{*}(t)$%
, equal to the re-summed approximant: 
\begin{equation}
\label{eq9}\Phi_2^{*}(t)=\phi_{21}^{*}(t)=1+a_1t\ \exp \left[ \frac{a_2}{a_1}%
t\right] ~. 
\end{equation}

We will also study, when necessary, the inverse function $\phi_{2I}(t)$ of $%
\phi_2(t)$, 
\begin{equation}
\label{eq10}\phi_{2I}(t)= 1/\phi_2(t) = 1-a_1t+(-a_2+a_1^2)\ t^2+...\equiv
1+b_1t+b_2t^2+... 
\end{equation}
The corresponding Froude number $F_I$ is 
\begin{equation}
\label{eq11}F_I \equiv {\frac{b_1^2 }{b_2}} = \frac{a_1^2}{a_1^2-a_2} = 
\frac{F_d}{F_d-1}~, ~~~~~~~ \left( F_d= \frac{F_I}{F_I-1}\right) ~. 
\end{equation}
In cases when better convergence can be achieved by working with the
``inverse'' series (\ref{eq10}), the observable $\Phi_2^{*}(t)$ will be
expressed through the approximant corresponding to the inverse series as
follows: 
\begin{equation}
\label{eq12}\Phi_2^{*}(t)=\left[ \phi _{21I}^{*}(t)\right]^{-1}=\left(
1+b_1t\ \exp \left[ \frac{b_2}{b_1}t\right] \right) ^{-1} 
\end{equation}

The convergence of our procedure is checked by estimating the value of the
multiplier \be
M_{21}(t)=\left( 1+\frac{a_2}{a_1}t\right) \exp \left( \frac{a_2}{a_1}t
\right) ~. \ee
The same expression gives the multiplier for the inversion function by
changing all the coefficients $a_i$'s into $b_i$'s, where the $b_i$'s are
related to the $a_i$'s via expression (\ref{eq10}).

\subsection{Negative velocity $a_1<0$ (downward trend)}

\subsubsection{Negative acceleration $a_2 < 0$}

Working with positive observables, this case corresponds to the possibility
that the observable vanishes in finite time. This zero-crossing occurs at $%
t_{c21}$ given by 
\begin{equation}
\label{jgnfda}1+F_dZ_d\exp \left[ Z_d\right] =0~, 
\end{equation}
where $Z_d={\frac{a_2}{a_1}}~t_{c21}$. The corresponding multiplier $%
M_{21}(t_{c21})$ is larger than $1$, signaling a possible problem with the
convergence of the functional renormalization method. This may signal an
instability in the time dynamics of the time series close to the
zero-crossing time.

\subsubsection{Moderate positive acceleration $0< a_2 < {\frac{a_1^2 }{e}}$}

The observable goes again to zero in finite time. The time $t_{c21}$ at
which the observable vanishes is given again by (\ref{jgnfda}). The
corresponding multiplier $M_1(t_{c21})$ is now less than $1$, signaling a
stable scenario.

\subsubsection{Strong positive acceleration ${\frac{a_1^2 }{e}} < a_2$ \label
{secmna}}

The acceleration is sufficiently positive to counter-balance the negative
trend and the observable goes through a minimum before rebounding upward.
The time $t_{{\rm min}}$ at which the minimum $\Phi_{\min }^{*}=1-\frac{F_d}{%
e}$ is reached is given by 
\begin{equation}
\label{fdafa}t_{{\rm min}}=-{\frac{a_1}{a_2}}~. 
\end{equation}
This is a very stable situation since the multiplier is zero at the minimum.
Our analysis suggests that the extrapolation to the future is the most
credible for this situation in the downward trend case.

\subsection{Positive velocity $a_1>0$ (upward trend)}

\subsubsection{Negative acceleration: $a_2<0$}

The observable increases up to a maximum $\Phi _{\max }^{*}=1-\frac{F_d}e$
and then decreases after it. The time $t_{{\rm max}}$ of the maximum is
given by 
$$
t_{{\rm max}}=-{\frac{a_1}{a_2}}~. 
$$
This is a very stable situation since the multiplier $M_{21}$ is less than $%
1 $ for arbitrary time $t$ and is zero at the maximum.

\subsubsection{Positive acceleration}

This case requires the inversion of the time series, since the multipliers
of the direct series are always larger than $1$: the expansion is not
convergent and does not supply any information on the real-axis
singularities. The inversion maps these two cases on the cases previously
considered with multipliers smaller than $1$. In addition, the corresponding
series has alternating signs, suggesting that finite-time singularity is
indeed present.

\paragraph{Moderate positive acceleration: $0<a_2<{\frac{e-1}e}~a_1^2$}

The observable increases up to a maximum and then decreases after it. The
time $t_{{\rm max}}$ of the maximum is given by 
\begin{equation}
\label{17}t_{{\rm max}}=-{\frac{b_1}{b_2}=\frac{a_1}{a_1^2-a_2}}~, 
\end{equation}
and the value at maximum is 
\begin{equation}
\label{18}\Phi _{\max }=\left( 1-\frac{F_I}e\right) ^{-1}~. 
\end{equation}
This solution is very stable and thus very credible as the multiplier for $%
\phi _{21I}^{*}(t)$ is always smaller than $1$ and vanishes at the maximum.
The original function is obtained by the inversion $\Phi _2^{*}(t)=\left(
\phi _{21I}^{*}(t)\right) ^{-1}$ (see also equation (\ref{eq10})).

\paragraph{Large positive acceleration: $\frac{e-1}{e}~ a_1^2 < a_2$ or $F_d
<\frac{e}{e-1}$}

There is an upward finite-time singularity at $t_{c21}$ given by the zero of
the inverted renormalized expansion, i.e., by the solution of 
\begin{equation}
\label{19}\phi _{21I}^{*}(t_{c21})=1+F_iZ_i\exp [Z_i]=0~, 
\end{equation}
{\rm where}~$~Z_i=t_{c21}~{\frac{a_2-a_1^2}{a_1}}$~. The multiplier for $%
\phi _{21I}^{*}(t)$ ) is smaller than $1$, signaling the convergence of the
theoretical procedure for the inverse functions. Recall again that, in order
to return to the original function, the inversion $\Phi _2^{*}(t)=\left(
\phi _{21I}^{*}(t)\right) ^{-1}$ should be performed, leading to a
divergence of the original observable at the point where its inverse crosses
zero.

\subsection{Example \label{example}}

Let us consider a function with known singular behavior and compare the
results obtained by means of self-similar approximations with the exact
values. We consider the function $1/\cos (t)$ which was found \cite{GSdamage}
to describe the time-evolution of the crack length of a self-consistent
model of damage and used \cite{GASpred} for prediction tests. Starting from $%
t=0$, the function $1/\cos (t)$ possesses a singularity at $t=\pi /2\approx
1.5708$, and its expansion up to second order in powers of $t^2$ reads 
\begin{equation}
\label{20}\phi _2(t)\simeq 1+1/2\ t^2+5/24\ t^4+... 
\end{equation}
Note that, since $1/\cos (t)$ is an even function, the relevant variable for
the expansion is indeed $t^2$ and our previous classification must be
applied to the expansion (\ref{20}) up to second order in $t^2$ (i.e., to
fourth-order $t^4$).

The corresponding Froude number is $F_d=1.2$ and obeys the condition $%
1<F_d<\frac e{e-1}$, i.e., the previous section shows that one can
anticipate a finite-time singularity, only on the basis of the quadratic
polynomial expansion. The expansion of the inverse of $\phi_2(t)$ reads 
\begin{equation}
\label{21}\phi_{2I}(t)\simeq 1-1/2\ t^2+1/24\ t^4+... 
\end{equation}

As expected from our previous analysis in terms of the stability of the
renormalization flow quantified by the multipliers, the position of the
critical time is better estimated from the zero of $\phi
_{21I}^{*}(t_{c21})=0$, which gives $t_{c21}=1.56645$, compared to the value 
$t_{20}=1.59245$ suggested from the condition $\phi _{20I}(t_{20})=0$. This
last value $t_{20}$ is nothing but the prediction of the critical time from
the ``bare'' polynomial expansion. We refer to (\ref{mgnglqw}) for the
definition of $\phi _{21I}^{*}(t)$ and $\phi _{20I}(t)$. This shows that
using a control parameter, which ensures that the renormalized expansion
retrieves the polynomial expansion, improves the prediction of the critical
time.

\section{Illustration of the selection of scenarios by the convergence of
higher-order approximants}

We now exploit the previous example of the function $1/\cos(t)$ described in
section \ref{example} to illustrate how the concept of convergence of the
approximants developed in section II can be used for an
improved determination of the singularity.

Higher-order expansions are given by

\begin{equation}
\label{22} \phi_{3I}(t)\simeq \phi_{2I}(t) -1/720\ t^6+...,~~~~~~
(b_3=-1/720) 
\end{equation}
and 
\begin{equation}
\label{23} \phi _{4I}(t)\simeq \phi_{3I}(t) +1/40320\ t^8+...,~~~~~~
(b_4=1/40320)~. 
\end{equation}
The corresponding higher order approximants are 
\begin{equation}
\label{24}\Phi _3^{*}(t)=\left( \phi _{3I}^{*}(t)\right) ^{-1}=\left(
1+b_1t^2\exp \left( \frac{b_2}{b_1}t^2\exp \left( \frac{b_3}{b_2}\tau
_2t^2\right) \right) \right) ^{-1}~, 
\end{equation}
with \be
\tau _2=1-\frac{b_2^2}{2b_1b_3}~, \ee
and 
\begin{equation}
\label{25}\Phi _4^{*}(t)=\left( \phi _{4I}^{*}(t)\right) ^{-1}=\left(
1+b_1t^2\exp \left( \frac{b_2}{b_1}t^2\exp \left( \frac{b_3}{b_2}\tau
_2t^2\exp \left( \frac{b_4}{b_3}\tau _3t^2\right) \right) \right)
\right)^{-1}~, 
\end{equation}
with 
\begin{equation}
\label{26}\tau _3=-\frac{b_3}{12b_1b_2b_4(b_2^2-2b_1b_3)}.\left(
24b_1^2b_4b_2+5b_2^4-12b_1^2b_3^2-12b_1b_3b_2^2\right)~. 
\end{equation}
As explained in section II, the control parameters $%
\tau_2 $ and $\tau_3$ are determined from the condition that the expansion
of the approximants at small $t$ coincide with the perturbative expression
of the initial polynomial representation.

For each approximant of a given different order and with a given number of
control parameters, we obtain a prediction for the position of the
singularity obtained from the condition of zero-crossing of the inverse
approximant. Let us compare these values between them and with the critical
time derived from the corresponding initial polynomial fit. We find the
following results: 
$$
~t_{c21}=1.56645,\qquad \qquad \quad \quad \quad \quad \qquad \qquad \qquad
\qquad \qquad \ \qquad 
$$
$$
t_{c31}=1.55134,\qquad \ t_{c32}=1.57067,\qquad \quad \quad \quad \quad
\qquad \qquad 
$$
$$
t_{c41}=1.55193,\ \qquad t_{c42}=1.57048,\qquad \ t_{c43}=1.57079; 
$$
corresponding to the following multipliers defined by (\ref{eq6}): 
$$
M_{21}(t_{c21})=0.6484,\qquad \qquad \quad \quad \quad \quad \qquad \qquad
\qquad \qquad \qquad \ \qquad 
$$
$$
M_{31}(t_{c31})=0.68955,\ M_{32}(t_{c32})=0.63708,\qquad \quad \quad \quad
\quad \qquad \qquad 
$$
$$
M_{41}(t_{c41})=0.68743,\ M_{42}(t_{c42})=0.63772,\ M_{43}(t_{c43})=0.63663. 
$$

The values $t_{c21}, t_{c31}, t_{c32}, t_{c41}, t_{c42}$ and $t_{c43}$
should be compared with the values determined from the ``bare'' polynomials 
$$
t_{20}=1.59245,~~~~t_{30}=1.56991,~~~~t_{40}=1.57082~. 
$$
The convergence of this last sequence can be analyzed by looking at the
corresponding sequence of multipliers $m_i(t,0)$ defined by (\ref{eq4}): 
$$
m_2(t_{02})=0.57735,~\ m_3(t_{03})=0.63985,~\ m_4(t_{04})=0.63651. 
$$

Overall these results show indeed a good convergence for $t_{c41},t_{c42}$
and $t_{c43}$ by increasing the number of control parameters and along the
diagonal $t_{c21},t_{c32},t_{c43}$ by increasing simultaneously both the
order of the polynomial and the number of control parameters. The same
observation holds for the multipliers $M$. In contrast, the bare polynomials
give a slower non-monotonous convergence. This confirms the improvement
brought by our scheme to obtain a better determination of the critical time $%
t_c$.

\section{Non-local control through the moments of the function to predict}

Section II and our subsequent tests have shown that the
control parameters provide a mean to improve the extrapolation to the future
by imposing some constraint on the approximants. Up to now, we have used the
constraint that the approximants must retrieve the ``bare'' polynomial
expansions at small times $t$. This corresponds to constraints which are
local in time.

It is interesting and potentially useful to investigate the possibility of
using more global ``non-perturbative'' constraints. A possible example is
when, either from a priori theoretical knowledge or from experimental or
empirical measurements, we get hold of the first $j-1$ moments $%
\mu_i,i=1,2..,j-1$ of the sought function $\phi(t)$ in some interval $[0,T]$%
: 
\begin{equation}
\label{27} \mu _i=\int_0^Tt^{i-1}\phi(t)dt~. 
\end{equation}
Our notation means that, for $j=2$, we know only the zero moment (``mass''
or integral $\phi(t)$ from $0$ to $T$), for $j=3$, we know both the zeroth
and first moment, etc...

Endowed with this knowledge of the first $j-1$ moments, we can condition the
control parameters $\tau _{1,}\tau _{2,..}\tau _{j-1}$ demanding that the
reconstructed approximants have exactly the right values of their moments: 
\begin{equation}
\label{28}\int_0^T\phi _j^{*}(t,\tau _{1,}\tau _{2,}...,\tau
_{j-1})~t^{i-1}dt=\mu _i~. 
\end{equation}
For $j=2$, we have one equation for $\tau _1$: 
$$
\int_0^T\phi _2^{*}(t,\tau _1)dt=\mu _0~. 
$$
For $j=3$, we obtain two equations for $\tau _1$ and $\tau _2$: 
$$
\int_0^T\phi _3^{*}(t,\tau _{1,}\tau _2)dt=\mu _0,\qquad \int_0^T\phi
_3^{*}(t,\tau _{1,}\tau _{2,})~t~dt=\mu _1~. 
$$
For $j=4$, we have three equations for $\tau _1,\tau _2$ and $\tau _3$: 
$$
\int_0^T\phi _4^{*}(t,\tau _{1,}\tau _2,\tau _3)dt=\mu _0~,\qquad
\int_0^T\phi _4^{*}(t,\tau _{1,}\tau _{2,}\tau _3)~t~dt=\mu _1~,\qquad
\int_0^T\phi _4^{*}(t,\tau _{1,}\tau _{2,}\tau _3)~t^2~dt=\mu _2~. 
$$

Based on these conditions, two different problems seem most natural. The
first one is to construct an approximate representation of the function $%
\phi(t)$ in the same interval $[0,T]$ where moments are given or measured.
In the case where the moments are obtained through some experimental
procedure leading to some measurement errors, this first problem amounts to
filter out the noise in the measurement interval $[0,T]$. The second problem
that we shall address here consists in extrapolating to times $t>T$.

Using the previous example of the function $1/\cos (t)$, let us consider the
following case $T=\sqrt{2}$. This value is ``natural'' as it is the root of $%
1+b_1t^2$. Constraining the control parameters by the knowledge of the
moments in the interval $[0,\sqrt{2}]$, we obtain the corresponding
approximants. The analysis of the zero of the inverse approximants give the
following estimations for the critical times and the corresponding
multipliers 
$$
t_{c21}=1.56888,\qquad t_{c32}=1.57077,\qquad t_{c43}=1.570796,\qquad 
$$
$$
M_{21}(t_{c21})=0.64388,\qquad M_{32}(t_{c32})=0.63678,\qquad
M_{43}(t_{c43})=0.63662~. 
$$
Notice the extremely good quality of the convergence of both the critical
times and of the multipliers.

It is also possible to use an hybrid approach, where some control parameters
are obtained from the agreement with the polynomial expansion at small time $%
t$, while the remaining ones are determined from the conditions on the known
moments. As an illustration, we show the fifth-order approximant 
\begin{equation}
\label{29}\phi _{5I}^{*}(t,\tau _1,\tau _{2,}\tau _3,\tau _4)=\left(
1+b_1t^2\exp \left( \frac{b_2}{b_1}\tau _1t^2\exp \left( \frac{b_3}{b_2}\tau
_2t^2\exp \left( \frac{b_4}{b_3}\tau _3t^2\exp (\frac{b_5}{b_4}\tau
_4t^2)\right) \right) \right) \right) ~, 
\end{equation}
where $\tau _1=1$ is conditioned by the polynomial expansion and the other
control parameters should be calculated from the system of equations \ba
\int_0^T\phi _{5I}^{*}(t,1,\tau _{2,}\tau _3,\tau _4)dt = \mu _0~, 
\nonumber 
\\ \int_0^T\phi _{5I}^{*}(t,1,\tau _{2,}\tau _3,\tau _4)tdt = \mu _1~, 
\nonumber
\\ \int_0^T\phi _{5I}^{*}(t,1,\tau _{2,}\tau _3,\tau _4)t^2dt = \mu _2~. 
\ea

Note that we do not even need to know the exact values $b_3,b_4$ and $b_5$
of the polynomial expansion since they can be included in the corresponding
controls. This results from the fact that the constraints on the moments
overwhelm the initial information on the coefficients of the polynomial
expansion. We find $t_{c54}=1.570796$ in extremely good agreement with the
exact critical time $t_c=\pi /2=1.5707963$.

\section{Classification and forecasting of critical times beyond mean-field}

We now use the formalism of section II and relax the
condition $s\to +\infty $ on the exponent of the algebraic transformation,
which amounted to impose the mean-field value $z=-1$ of the critical
exponent. The control exponent $s$ will be determined from the optimization
of the convergence and the stability of the renormalization flow. according
the general principles developed by Yukalov and Gluzman \cite{G6}-\cite{G13}.

\subsection{General procedure \label{genprofj}}

Consider as before an expansion of an observable $\phi (t)$ in powers of a
variable $t$ (time) given by $\phi _k(t)=\sum_{n=0}^ka_n~t^n$ where $a_0=1$
without loss of generality by suitable normalization. The method of
algebraic self-similar renormalization \cite{G6}-\cite{G8} gives the
following general recurrence formula for the approximant of order $k$ as a
function of the expansion $\phi _{k-1}(t)$ up to order $k-1$: 
\begin{equation}
\label{30}\phi _k^{*}(t)=\phi _{k-1}(t)\left[ 1-\frac{k~a_k}s~t^k~\phi
_{k-1}^{k/s}(t)\right] ^{-s/k}\equiv \left[ \phi _{k-1}^{-k/s}(t)-\frac{k~a_k%
}s~t^k\right] ^{-s/k}~, 
\end{equation}
where, in general, $s=s_k(t)$ depends on the approximation number $k$ and on
the variable $t$. These approximants automatically agree with their
corresponding polynomial expansions and the sole way to impose some control
is to restrict $s$ using some conditions of rather general nature such as
convergence of the sequence of approximants.

In the sequel, we assume that only the second-order expansion is available.
In order to determine the critical exponent $z$, we follow Yukalov and
Gluzman \cite{G8} and construct the two approximants available from the
knowledge of the two coefficients $a_1$ and $a_2$. They can be readily
obtained from the general formula (\ref{30}). The first order approximant is
simply 
\begin{equation}
\label{31} \Psi_1^{*}(t)=\left( 1-{\frac{a_1}{s_1}}t\right)^{-s_1} ~. 
\end{equation}
Representing $\phi_2(t)$ as $\phi_2(t)=1+a_1t(1+\frac{a_2}{a_1}~t)$ and
applying the general formula to the expression in brackets, we obtain the
second order approximant 
\begin{equation}
\label{32} \Psi_2^{*}(t)=1+a_1t\left( 1-{\frac{a_2}{a_1s_2}}t\right)^{-s_2}
~. 
\end{equation}
Let us assume further that $s_{1}=s_2=s$, where $s$ is the limiting value of
the control function of the algebraic transformation at the critical point.
It is apparent from the form of (\ref{30}) that $s$ plays the role of the
critical index $z$. As it was explained in \cite{G8}, this is justified in
the vicinity of a stable fixed point.

The condition of maximum stability of the renormalization flow is equivalent
to imposing that the difference $\Psi _2^{*}-\Psi _1^{*}$ be a minimum with
respect to the set of parameters (the so-called minimal difference
condition). We discuss below an application of the technique applied to
direct second-order expansions.

In the present work, we are interested in testing for the existence of a
finite-time singularity or critical point. Looking for such an occurrence,
we need to solve the minimal difference condition which amounts to look for
the solutions of the two equations in terms of the two variables $t_c$ and $%
s $: 
\begin{equation}
\label{33}\Psi _1^{*}(t_c,s)=0~~~~~~{\rm and}~~~~~~\Psi _2^{*}(t_c,s)=0~. 
\end{equation}
The vanishing of $\Psi _1^{*}$ given by (\ref{31}) gives 
\begin{equation}
\label{34}t_c=s/a_1~. 
\end{equation}
The second condition $\Psi _2^{*}=0$ with (\ref{32}) provides an estimate $s$
for the critical index. In terms of Froude parameter $F_d$ defined by (\ref
{eq8}), the second condition of (\ref{33}) can be conveniently written as 
\begin{equation}
\label{35}1+s\ \left( 1-F_d^{-1}\right) ^{-s}=0~. 
\end{equation}
This gives $s=s(F_d)$ as a function of Froude number. In the cases when this
equation does not have a real solution, we determine the control parameter $%
s $ (which, we recall, is more general entity than the critical index) from
the minimization of $\Psi _2^{*}(t_c,s)$: 
\begin{equation}
\label{36}\min _s\left( 1+s\ \left( 1-F_d^{-1}\right) ^{-s}\right) ~. 
\end{equation}

The Yukalov-Gluzman technique then confronts the two approximants $%
\Psi_1^{*} $ and $\Psi_2^{*}$: after their difference is minimized, it
remains to decide which one of them is the best re-summed expression
originating from the original perturbative expansion. Implicit in this
approach is the concept that the renormalization approach might not be fully
convergent asymptotically but only locally. Such a decision can be made
based on the analysis of the corresponding multipliers

$$
M_j(t,s)\equiv [\frac{\partial \Psi _j^{*}(\varphi ,s)}{\partial \varphi }%
]_{_{\varphi =P_1(t,0)}},~~~~~~~~~~j=1,~2~, 
$$
yielding

\begin{equation}
\label{37a}M_1(t,s)=\left( 1-\frac{a_1t}s\right) ^{-\left( 1+s\right) }, 
\end{equation}
\begin{equation}
\label{37b}M_2(t,s)=\left( 1-\frac{a_2}{a_1s}\ t\right) ^{-s}\left[ 1+\frac{%
a_2}{a_1}t\left( 1-\frac{a_2}{a_1s}\ t\right) ^{-1}\right] ~. 
\end{equation}

The more stable solution corresponding to the smallest $\left| M_j\right| \ $%
should then be selected. We find in general that $\Psi _1^{*}$ has the
smallest multiplier in the critical region, which onset is determined by the
condition $\left| M_1(t,s)\right| <<\left| M_2(t,s)\right| $, provided that
a solution to (\ref{35}) or (\ref{36}) exists. On the other hand, we find
that $\Psi _2^{*}\ $prevails in some ``pseudocritical'' regime when the
first solution $\Psi _1^{*}\ $becomes unstable. One can make this selection
process automatic by the weighting procedure advocated in \cite{G13} which
has also been used in \cite{Techanal}. The weighting procedure amounts to
defining an average of the two approximants with weights inversely
proportional to their multipliers. The rational for this approach is that
the inverse of the multipliers can be shown to play a role similar to the
probability that the system visits the dynamical state described by the
corresponding approximant. The resulting function is 
\begin{equation}
\label{38}\Psi ^{*}(t,s)=\frac{\Psi _1^{*}(t,s)\left| M_1(t,s)\right|
^{-1}+\Psi _2^{*}(t,s)\left| M_2(t,s)\right| ^{-1}}{\left| M_1(t,s)\right|
^{-1}+\left| M_2(t,s)\right| ^{-1}}~. 
\end{equation}

Since we are interested in the physically meaningful case when the critical
time $t_c=s/a_1$ is positive (it may be infinite), a modification is
required when $s/a_1$ is found negative or simply where there are no real
solution to (\ref{33}). We take this situation as a signal that one should
use the inverse function defined by (\ref{eq10}) as the relevant expansion
to obtain the most stable scenario. The sought function is then defined as
the inverse of the weighted-average for inverse renormalized approximant.
The approximants and corresponding multipliers are calculated using the
parameters $a_i$'s (equation (\ref{eq7})) changed into $b_i$'s (equation (%
\ref{eq10})). The final solution reads 
\begin{equation}
\label{39}\Psi^{*}(t,s)\equiv \Psi _I^{*}(t,s)^{-1}=\left( \frac{\Psi
_{1I}^{*}(t,s)\left| M_{1I}(t,s)\right| ^{-1}+\Psi _{2I}^{*}(t,s)\left|
M_{2I}(t,s)\right| ^{-1}}{\left| M_{1I}(t,s)\right| ^{-1}+\left|
M_{2I}(t,s)\right| ^{-1}}\right) ^{-1}~. 
\end{equation}

\subsection{Negative velocity $a_1<0$ (downward trend)}

Using the general procedure of section \ref{genprofj}, we now present the
corresponding classification of the different possible regimes.

\subsubsection{Negative acceleration $a_2<0$ \label{jbgfq}}

In this situation, the inverse function has always a singularity. This
corresponds for the direct observable function to vanish in finite time
(critical regime I) at $t_c=s(F_d)/a_1$, where $s(F_d)$ is the negative
solution of equation (\ref{35}). Both approximants $\Psi _1^{*}$ and $\Psi
_2^{*}$ contribute to the expression (\ref{39}) and $\Psi _2^{*}$
progressively dominates at time approaches $t_c$.

Solution $\Psi _2^{*}$ contribute to the average (\ref{38}) more than $\Psi
_1^{*}$ , because $\left| M_2\right| $ always smaller than$\left| M_1\right|
.$ As the multiplier of $\Psi _1^{*}(t)$ eventually blows up to infinity at $%
t_c$, $\Psi _2^{*}(t)$ ends by dominating the behavior of $\Psi ^{*}(t)$ .

The asymptotic behavior of the average close to $t_c$ is determined by $\Psi
_2^{*}$ and is characterized by exponent $z=1$, notwithstanding the fact
that the control exponent $s$ is fractional. This corresponds to the
situation where the observable goes to zero linearly in time.

\subsubsection{Positive moderate acceleration $0<a_2<\frac{a_1^2}{F_0}$, $%
F_0 < F_d$ where $F_0=\left( 1-e^{-1/e}\right)^{-1}$}

In this region of parameter $F_0=\left( 1-e^{-1/e}\right) ^{-1}=3.249<F_d$,
the observable still goes to zero in finite time (critical regime I). The
time $t_c$ at which the observable vanishes is given by the same formula as
in section \ref{jbgfq}. This solution exists as long as there is solution $%
s(F_d)$ to (\ref{35}). When $F_d$ becomes too small, (\ref{35}) has no more
any solution and this corresponds to the critical regime II discussed in the
next section. The boundary between these two regimes occurs at the Froude
value $F_0$ determined by adding the condition 
\begin{equation}
\label{40}\frac{\partial (\Psi _2^{*}(s/a_1,s))}{\partial s}=0 
\end{equation}
to the general equation $\Psi _2^{*}(s/a_1,s)=0$. The minimum, solution of (%
\ref{40}), is located at 
\begin{equation}
\label{41}s_{\min }=\frac 1{\ln \left( \frac{F_d-1}{F_d}\right) }, 
\end{equation}
and coincides with the zero of $\Psi _2^{*}$ only for the specific value of
the Froude number $F_0$ thus determined by 
\begin{equation}
\label{42}F_0=\left( 1-e^{-1/e}\right) ^{-1}=3.249~. 
\end{equation}
As $F_d\rightarrow \infty $, equation (\ref{35}) can be solved exactly and $%
s(F_d\rightarrow \infty )=-1,$ which is a mean-field value $z=1$. In other
words, given in advance a linear function $\phi (t)=1-\left| a_1\right| t,$
our technique will indeed reconstruct it! Let us expand Eq.(35) around this
exactly solvable limit in powers of a small parameter $1/F_d=y$,\be
1+s\ \left( 1-F_d^{-1}\right) ^{-s}\simeq 1+s+s^2y+.... \ee
Then, \be
s=\frac 1{2y}\left( -1+\sqrt{1-4y}\right) \simeq -1-y+...\ \left(
y\rightarrow 0\right) . \ee
This expression breaks down around $F_d=4,$ where the whole equation (\ref
{35}) should be considered down to $F_0,$where $s(F_0)=-e.$

Solution $\Psi_1^{*}$ starts to contribute to the average (\ref{38}) more
than $\Psi_2^{*}$, as soon as $t$ satisfy condition $\left| M_1\right| \ll
\left| M_2\right|$, as shown in Fig.~1, which represents the dependence of
the two approximants $\Psi _1^{*}(t)$ (equation (\ref{31})) and $%
\Psi_2^{*}(t)$ (equation (\ref{32})) and of their weighted average $\Psi
^{*}(t)$ given by (\ref{38}). As the multiplier of $\Psi _1^{*}(t)$
eventually vanishes at $t_c$, $\Psi _1^{*}(t)$ ends by dominating the
behavior of $\Psi ^{*}(t)$ and the average demonstrates critical behavior
with positive fractional exponent $z=-s(F_d)$. Thus, in this region of $F_d$%
, as $t$ goes to $t_c$, we obtain a critical behavior with fractional $s$
playing the role of critical index $z$. It means that the exponent is now
different from $-1$ and is determined by the solution of eq.~(\ref{35}).

\subsubsection{Strong positive acceleration ${\frac{a_1^2}{{F_0}}}<a_2\ $, $%
F_d < F_0$}

The critical regime (I) is now replaced by the critical regime (II), such
that equation (\ref{35}) has no more any solution and the control exponent
and critical time are determined from the minimization of eq.~(\ref{36}),
which gives \be
s=s_{\min }(F_d)=\frac 1{\ln \left( \frac{F_d-1}{F_d}\right) } ~. \ee
The critical index is $z=-s_{\min }(F_d)$, leading to a logarithmic
correction to the mean-field value. The critical time is given by 
\begin{equation}
\label{43}{\rm ~~~}t_c=\frac{s_{\min }(F_d)}{a_1}~. 
\end{equation}
Figure 2 shows the dependence of the two approximants $\Psi _1^{*}(t)$
(equation (\ref{31})) and $\Psi _2^{*}(t)$ (equation (\ref{32})) and of
their weighted average $\Psi ^{*}(t)$ given by (\ref{38}). The new feature
is the existence of a minimum at time $t_{{\rm min}}$ of $\Psi _2^{*}(t)$
and, therefore of a non-monotonous behavior also of the average $\Psi
^{*}(t),$ with $t_{{\rm min}}$ given by 
\begin{equation}
\label{44}t_{{\rm min}}=-{\frac{a_1}{a_2}\frac{s_{\min }}{{s_{\min }-1}}}~, 
\end{equation}
corrected by the ratio $\frac{s_{\min }}{{s_{\min }-1}}$ compared to the
mean-field result (\ref{fdafa}) of section \ref{secmna}. The value of $\Psi
_2^{*}(t)$ at $t_{{\rm min}}$ is 
$$
\Psi _{2\min }^{*}=1-F_d\left( 1+L(F_d)\right) ^{-\left( 1+L(F_d)\right)
/L(F_d)}~,\quad \quad L(F_d)=\ln \left( \frac{F_d}{F_d-1}\right) ~. 
$$
The trajectory $\Psi ^{*}(t)$ shown in figure 2 is rather unusual since
after spending some time close to the local minimum of a non-critical branch 
$\Psi _2^{*}(t)$, the system suddenly breaks down towards the ``critical''
branch $\Psi _1^{*}(t)$, which then ends at a critical point ${\rm ~}t_c$.
This means that the critical behavior with exponent $z=-s_{\min }(F_d)$ has
not disappeared yet. The drop occurs at a crossover time $t=t_{cros}$
defined as the solution to the equation $\left| M_1\right| \approx \left|
M_2\right| $, with a magnitude $\Delta =\Psi _2^{*}(t_{cros})-\Psi
_1^{*}(t_{cros})$. This regime is found for the Froude interval $%
F_{01}<F_d<F_0$ where $F_{01}=\left( 1-e^{-1}\right) ^{-1}=1.582$ is the
solution of the equation 
\begin{equation}
\label{46}1+\frac 1{\ln (\frac{F_{01}-1}{F_{01}})}=0~, 
\end{equation}
corresponding to the Froude value at which the multiplier $M_1(t,s)$ changes
from stable ($M_1<1$) to unstable ($M_1>1$) behavior. As the regime $%
1<F_d<F_{01}$ (pseudocritical regime I) sets in, the multiplier $M_1(t,s)$
becomes larger than $1$, increases with time and diverges at $t_c.$ Rather
than converging to the $\Psi_1^{*}(t)$ approximant, the weighted average $%
\Psi ^{*}(t)$ exhibits a fast change of direction to reach $\Psi _2^{*}(t)$
at $t_c$, as shown in figure 3. The critical branch has disappeared as the
non-critical branch $\Psi _2^{*}$ dominates. The presence of the approximant
scenario $\Psi _1^{*}$ is felt only in the existence of some oscillations
decorating the $\Psi _2^{*}$ scenario. This regime exists for $1<F_d<F_{01}$.

For $F_d<1\ $( pseudocritical regime II), the minimum of (\ref{36})
disappears and there is no solution either to (\ref{35}) or to (\ref{36})
anymore. This implies that we should use the inverse expansion in terms of
the coefficients $b_1>0,b_2<0\ \left( F_I<0\right) $. The corresponding
control exponent $s$ is then obtained from the condition 
\begin{equation}
\label{47}1+s\ \left( 1-F_I^{-1}\right) ^{-s}=0~, 
\end{equation}
which has a negative solution $s=s(F_I)$ leading to $t_c=s/b_1<0$, which is
not allowed. However, similarly to the previous strategy to replace (\ref{35}%
) by (\ref{36}), we can look for the solution which minimizes the
left-hand-side of (\ref{47}), which gives $s\rightarrow \infty $. This
corresponds to a pseudocritical regime which is reminiscent of the last
phase of the previous regime, but with formally infinite $t_c.$ In the limit 
$s\rightarrow \infty $, we obtain 
$$
\Psi _{1I}^{*}(t)=\exp (b_1t),\quad \quad M_{1I}(t)=\exp (b_1t) 
$$
$$
\Psi _{2I}^{*}(t)=1+b_1t\ \exp (\frac{b_2}{b_1}t),\quad M_{2I}(t)=\left( 1+ 
\frac{b_2}{b_1}t\right) \exp \left( \frac{b_2}{b_1}t\right) . 
$$
Note that $\Psi_{2I}^{*}(t)$ is qualitatively similar to the mean-field
solution of section III, derived for the same region of
parameters. The multiplier $M_{1I}(t)$ is always larger than $1$ which
implies that the scenario $\Psi_{2I}^{*}$ always dominates in the weighted
average, although the contribution of $\Psi_{1I}^{*}$ is responsible for the
existence of an extra- minimum in the trajectory of $\Psi ^{*}$ given by
expression (\ref{39}), as shown in figure 4.

\subsection{Positive velocity $a_1>0$ (upward trend)}

\subsubsection{Negative acceleration: $a_2<0$}

One can still define the control exponent $s$ from the condition (\ref{35}),
but $t_c$ becomes negative which is undesirable. As in the previous section,
we turn to the next possibility which is to use (\ref{36}), whose only
solution is $s\rightarrow +\infty $. This solution turns out to minimize the
difference between two approximant scenarios. This regime is the inversion
of the pseudocritical regime II just presented above (only with $F_d<0$
instead of $F_I$), as it is described by the following solutions, 
\begin{equation}
\label{49}\Psi _1^{*}(t)=\exp (a_1t)~,\quad \quad M_1(t)=\exp (a_1t)~. 
\end{equation}
\begin{equation}
\label{50}\Psi _2^{*}(t)=1+a_1t\ \exp (\frac{a_2}{a_1}t),\quad M_2(t)=\left(
1+\frac{a_2}{a_1}t\right) \exp \left( \frac{a_2}{a_1}t\right) ~. 
\end{equation}
Note that $\Psi _2^{*}(t)$ is nothing but the mean-field solution of section 
III, derived for the same parameter region. The multiplier $M_1(t)$
is found to be always larger than $1$. Therefore, $\Psi _2^{*}$ dominates in
the weighted average trajectory $\Psi ^{*}$(\ref{38}). The contribution from 
$\Psi _1^{*}$ induces splitting of the mean-field maximum (at $t=-a_1/a_2$)
as shown in figure 5.

\subsubsection{Moderate positive acceleration: $0<a_2<{\frac{{F_0}-1}{{F_0}}}
~a_1^2$; $\frac{F_0}{F_0-1}< F_d$}

In this case, although equation (\ref{35}) has a solution for $s<0$, the
corresponding $t_c$ is negative. After inversion of the initial series, this
region of parameters is equivalent to $1<F_I<F_0$. This regime corresponds
to the inverse of the critical regime II described above and can be
described similarly.

Consider first the region of $F_{01}<F_I<F_0$.\ There is a minimum of the
curve $\Psi _{2I}^{*}$ and the shape of the observable $\left( \Psi
_I^{*}\right) ^{-1}$ (see equation (\ref{39})) is significantly
non-monotonous, due to contribution from $\Psi _{2I}^{*}$. The time $t_{{\rm %
min}}$ of the minimum of $\left( \Psi _{2I}^{*}\right) ^{-1}$ (maximum of $%
\Psi _{2I}^{*})\ $ is given by 
\begin{equation}
\label{52}t_{{\rm min}}=-{\frac{b_1}{b_2}\frac{s_{\min }}{{s_{\min }-1}}~,} 
\end{equation}
where 
\begin{equation}
\label{53}s_{\min }=\frac 1{\ln \left( \frac{F_I-1}{F_I}\right) } 
\end{equation}
is the solution of the minimization 
$$
\min _s\left( 1+s\ \left( 1-F_I^{-1}\right) ^{-s}\right) ~. 
$$
The maximum value is $\left( \Psi _{2I(\min )}^{*}\right) ^{-1}$ where 
\begin{equation}
\label{54}\Psi _{2I(\min )}^{*}=\left( 1-F_I\ \left( 1+L(F_I)\right)
^{-\left( 1+L(F_I)\right) /L(F_I)}\right) ,\quad L(F_I)=\ln \left( \frac{F_I 
}{F_I-1}\right) ~. 
\end{equation}
The trajectory has the same topology as shown in figure 2 except that it is
the inverse of the function shown in figure 2. The observable is predicted
as the weighted average scenario given by (\ref{39}) and is shown in figure
6. The ``critical'' branch $\Psi _{1I}^{*}$ shapes the weighted average $%
\Psi _I^{*}$ as $t_c$ is approached. The weighted average scenario goes to
infinity in finite time $t_c=s_{\min }/b_1,$ with negative $z=s_{\min }$
describing the power-low divergence. In terms of the coefficients $a_i$ of
the polynomial expansion, this regime holds for $\frac{F_{01}-1}{F_{01}}%
~a_1^2<a_2<\frac{{F_0}-1}{F_0}~a_1^2$, i.e., for $\frac{F_0}{F_0-1}<F_d< 
\frac{F_{01}}{F_{01}-1}$ ($\frac{F_{01}-1}{F_{01}}=0.368,~\frac{F_0-1}{F_0}%
=0.692$). These conditions are equivalent to $F_{01}<F_I<F_0$.

As $F_I$ becomes smaller than $F_{01}$, the multiplier $M_{1I}$(t,s) changes
from stable ($M_{1I}<1$) to unstable ($M_{1I}>1$). As a consequence and
similarly to the change from figure 2 to figure 3, the weighted average
scenario changes dramatically and does not exhibit anymore a critical
divergence at $t_c$. The scenario $\Psi _{1I}^{*}$ is felt only in the
creation of a few oscillations around $\Psi _{2I}^{*}$. This regime holds
for $1<F_I<F_{01}$ and mirrors the pseudocritical regime I previously
described. In terms of initial coefficients, it corresponds to $0<a_2<{\frac{%
{F_{01}}-1}{{F_{01}}}}~a_1^2$.

\subsubsection{Large positive acceleration: ${\frac{{F_0}-1}{{F_0}}}
~a_1^2<a_2$, $F_I>F_0$}

In terms of the inverse Froude number, this regime holds for $F_I>F_0$. The
observable goes to infinity in finite time at a critical time $t_c$, which
is determined from the condition that the inverse quantities cross zero. The
corresponding control exponent $s(F_I)$ is the solution of 
\begin{equation}
\label{55}1+s\ \left( 1-F_I^{-1}\right) ^{-s}=0~, 
\end{equation}
and%
$$
t_c=s(F_I)/b_1. 
$$
The existence of the finite-time singularity holds as long as $F_I>F_0$.
This condition can be re-expressed in terms of the direct Froude number and
gives $F_d<\frac{F_0}{F_0-1}=1.445$. This regime mirrors the critical regime
I. Given a second-order expansion $\phi _2(t)\simeq 1+\left| a_1\right|
t+\left| a_1\right| ^2\ t^2\ $of a simple pole $\phi (t)=\left( 1-\left|
a_1\right| t\right) ^{-1}$ (with $F_I\rightarrow \infty $), our technique
will indeed reconstruct it! The weighted average scenario $\Psi
_I^{*}(t,s)^{-1}$goes to infinity in finite time $t_c,$ with negative $%
z=s(F_I)$ describing the power-low divergence. In Fig.~7 we demonstrate the
dependence of the two approximants $\Psi _1^{*}(t)$ and $\Psi _2^{*}(t)$ and
of their weighted average $\Psi ^{*}(t)$ given by (\ref{39}) .

\section{Concluding Remark}

Starting from a representation of the early time evolution of a dynamical
system in terms of the polynomial expression of some observable $\phi (t)$
as a function of time, we have investigated the conditions under which this
early time dynamics may lead or may not lead to a finite-time singularity.
The corresponding classification has been performed from the point of view
of the functional renormalization method of Yukalov and Gluzman \cite{G1}-%
\cite{G13}, with the purpose of identifying the most stable scenarios, given
the early time dynamics. The direct extension of this work is to test our
predictions empirically, following the methodology of \cite{GASpred}
developed for a particular case.

\pagebreak

{\bf Figures captions} \vskip 1cm

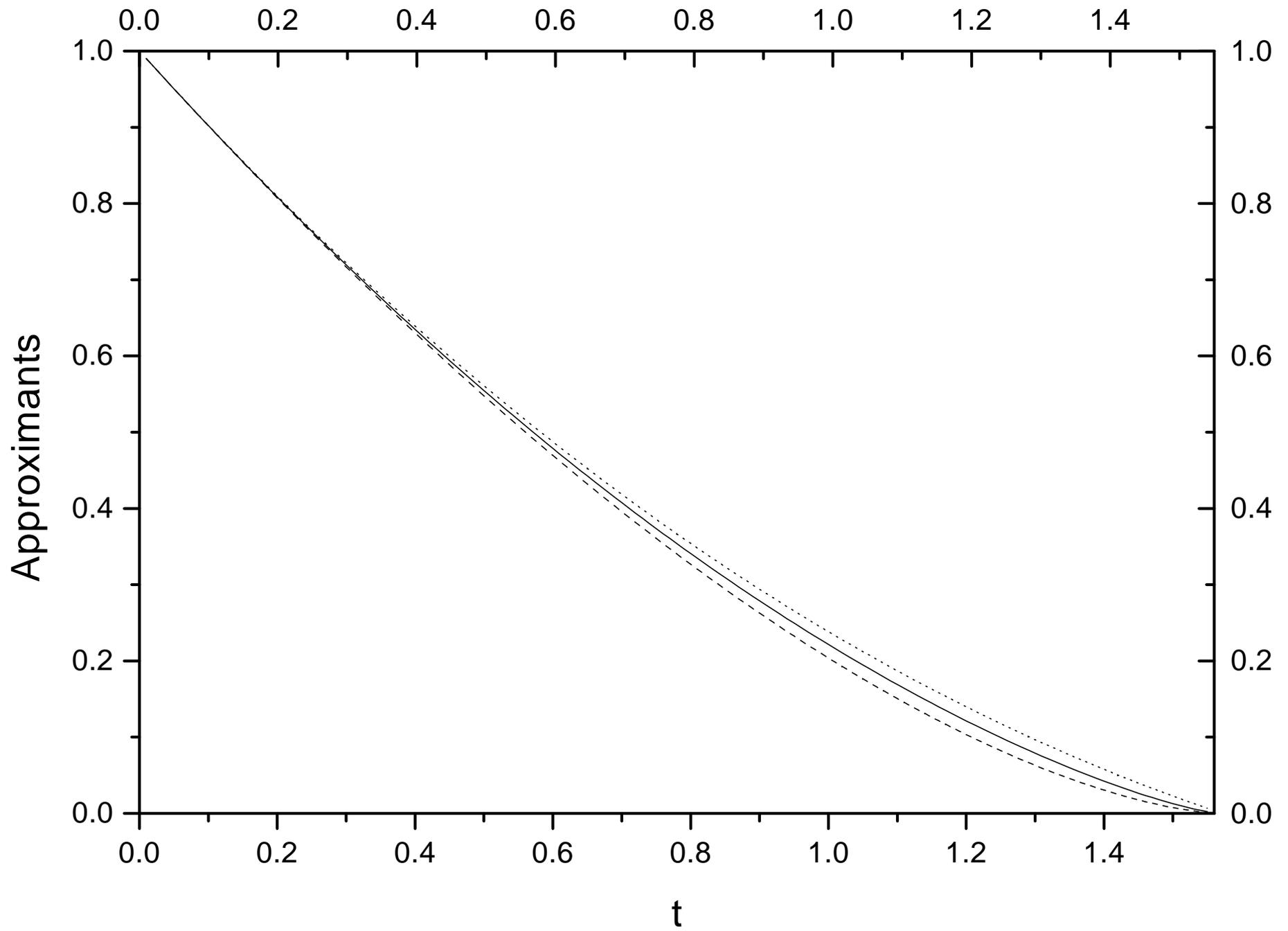
Fig.~1: First-order approximant $\Psi _1^{*}$ (equation (\ref{31})), dashed
line), second-order approximant $\Psi _2^{*}$ (equation (\ref{32}), dotted
line) and their weighted average given by (\ref{38}) (continuous line) as a
function of time, for positive moderate acceleration $0<a_2<\frac{a_1^2}{F_0}
$, $F_0<F_d$ where $F_0=\left( 1-e^{-1/e}\right) ^{-1}$.

\vskip 1cm

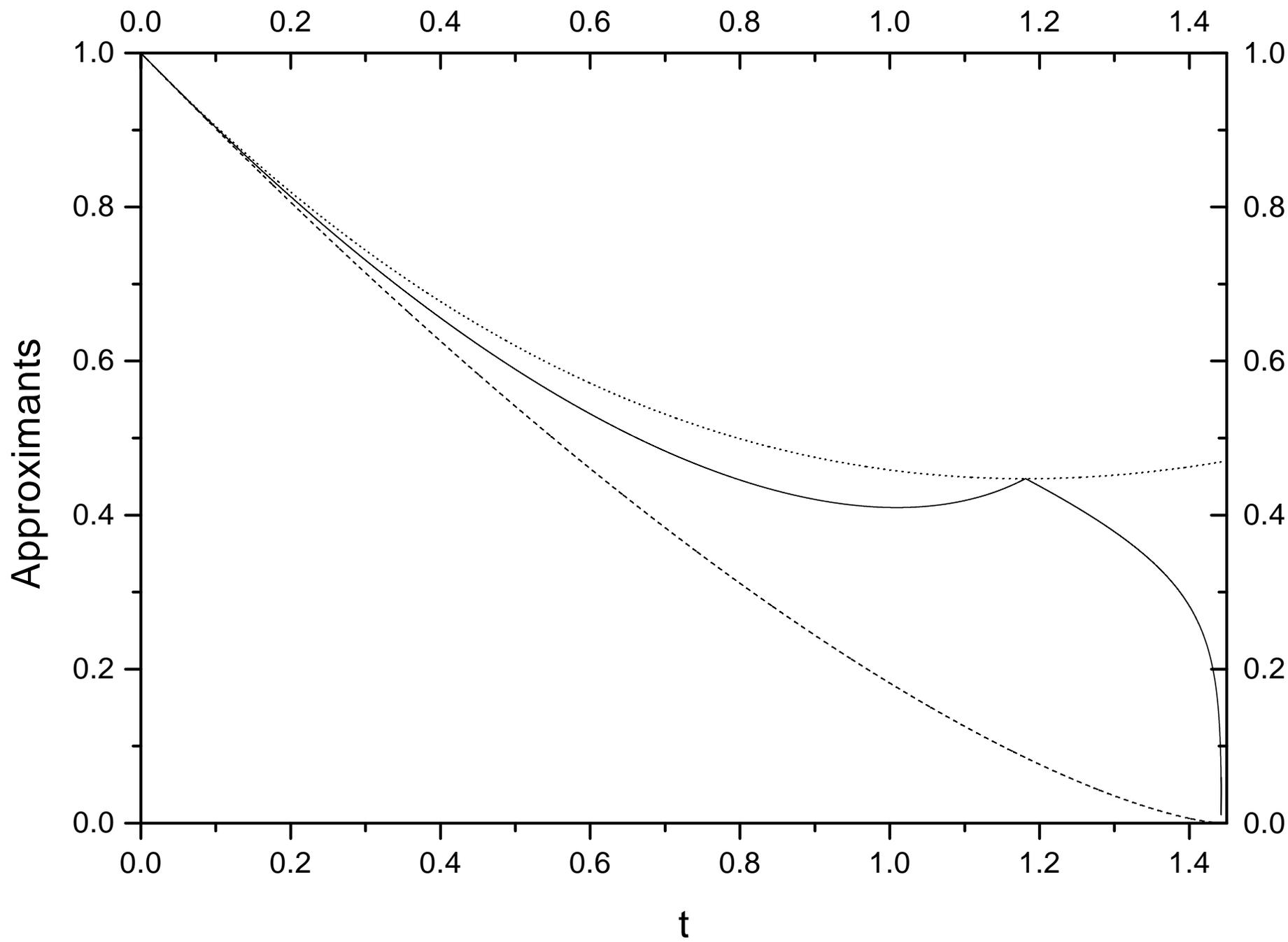
Fig.~2: Dependence of the two approximants $\Psi_1^{*}(t)$ (equation (\ref
{31}), dashed line) and $\Psi _2^{*}(t)$ (equation (\ref{32}), dotted line)
and of their weighted average $\Psi^{*}(t)$ given by (\ref{38}) (continuous
line) in the regime of strong positive acceleration $\frac{a_1^2}{F_0}<a_2$, 
$F_d < F_0$.

\vskip 1cm

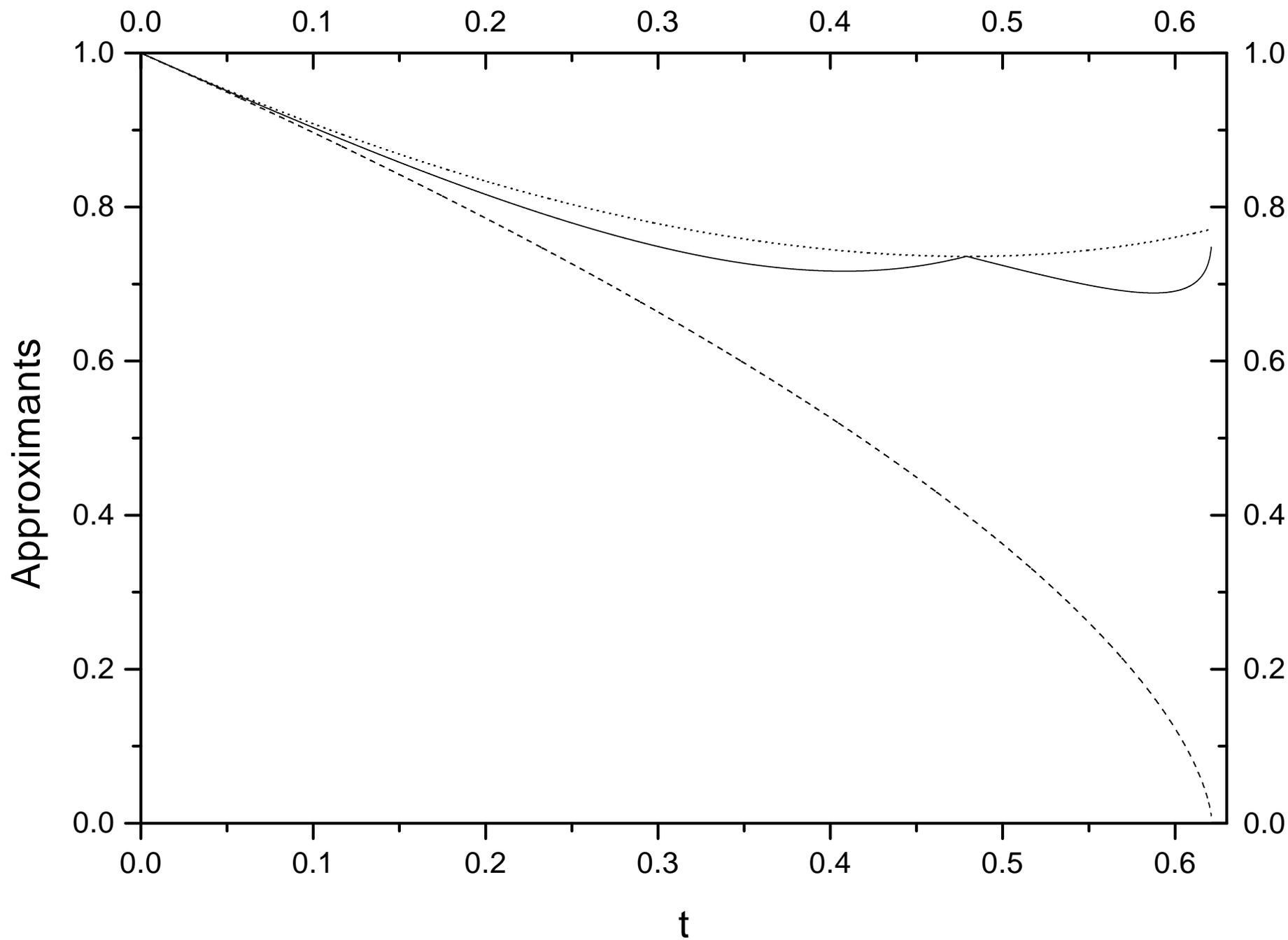
Fig.~3: Same as figure 2 in the regime $1<F_d<F_{01}$ (pseudocritical regime
I). Note that the weighted average $\Psi ^{*}(t)$ exhibits a fast change of
direction to reach $\Psi _2^{*}(t)$ at $t_c$.

\vskip 1cm

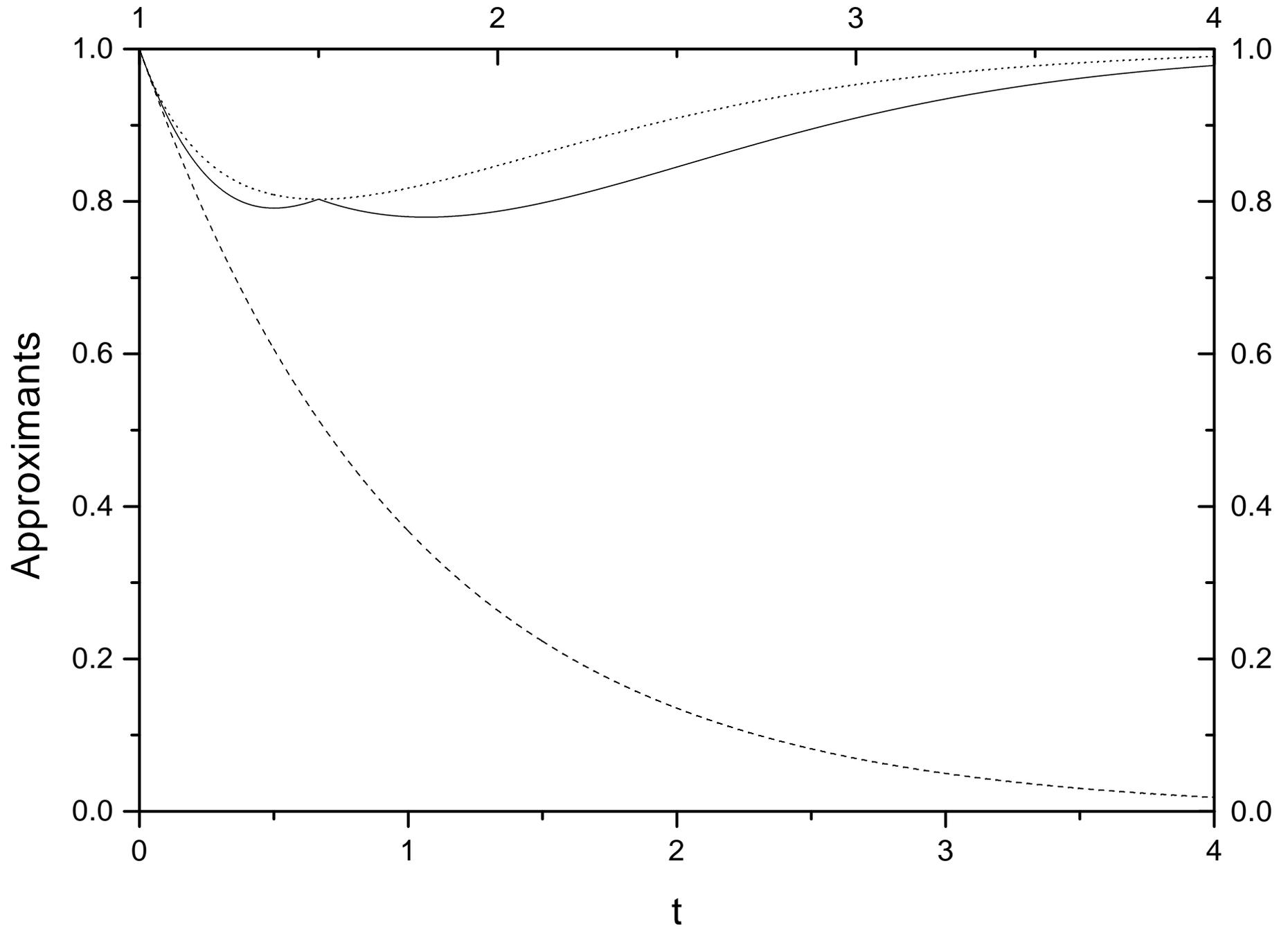
Fig.~4: Same as figure 2 for $F_d<1$.

\vskip 1cm

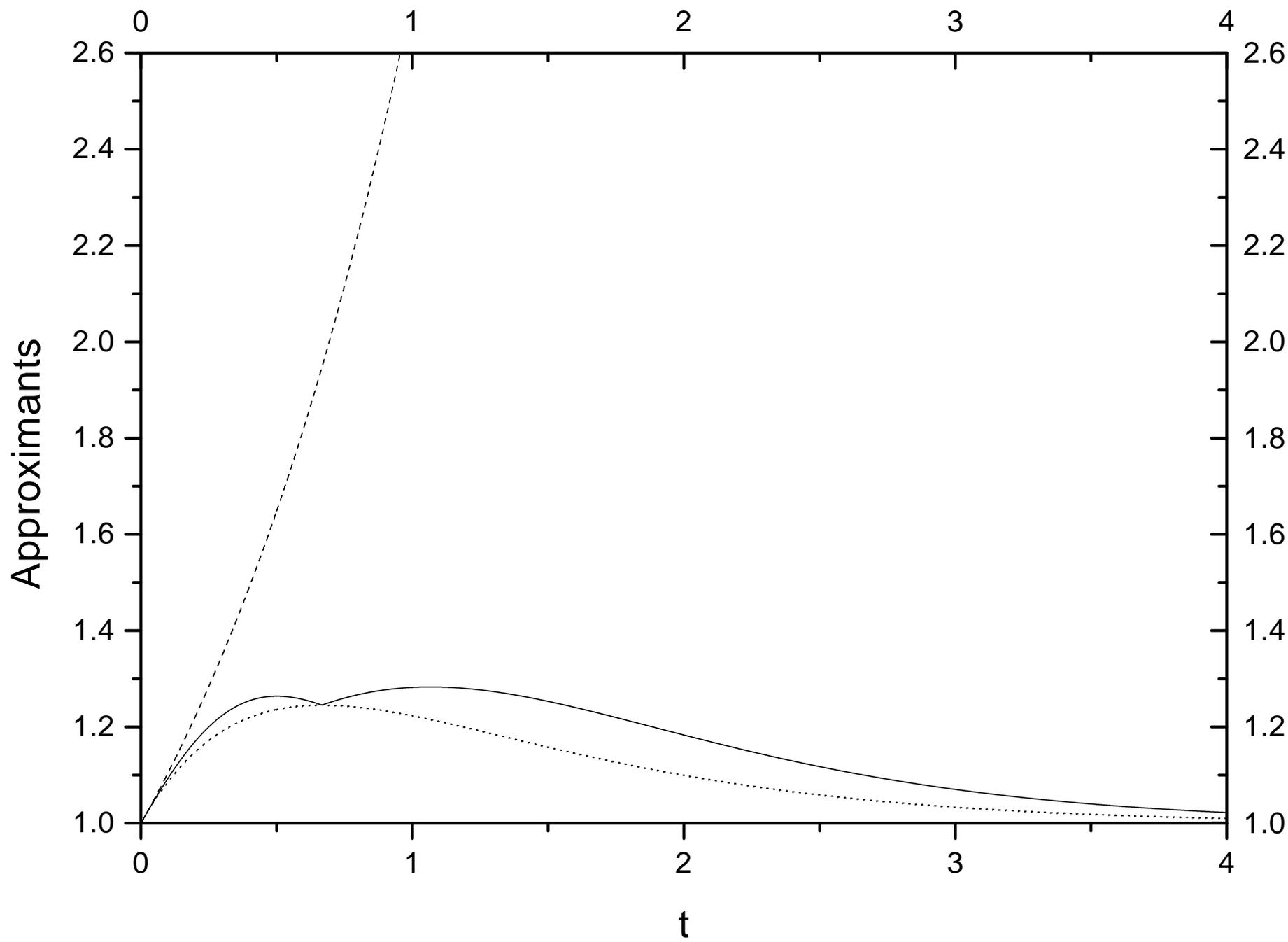
Fig.~5: Same as figure 2 for a negative acceleration $a_2<0$.

\vskip 1cm

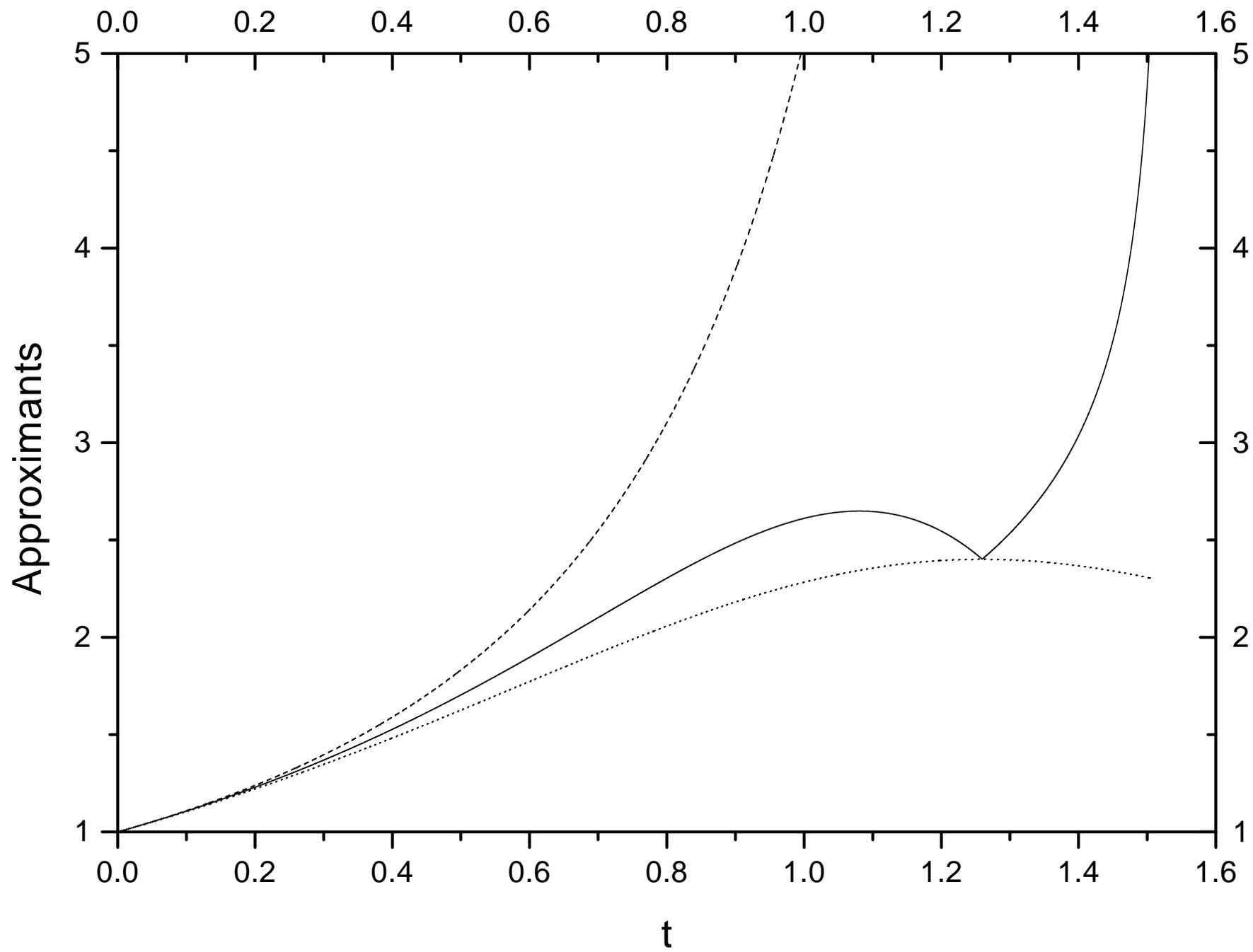
Fig.~6: Same as figure 2 in the regime of moderate positive acceleration: $%
0<a_2<{\frac{{F_0}-1}{{F_0}}}~a_1^2$; $\frac{F_0}{F_0-1}<F_d$.

\vskip 1cm

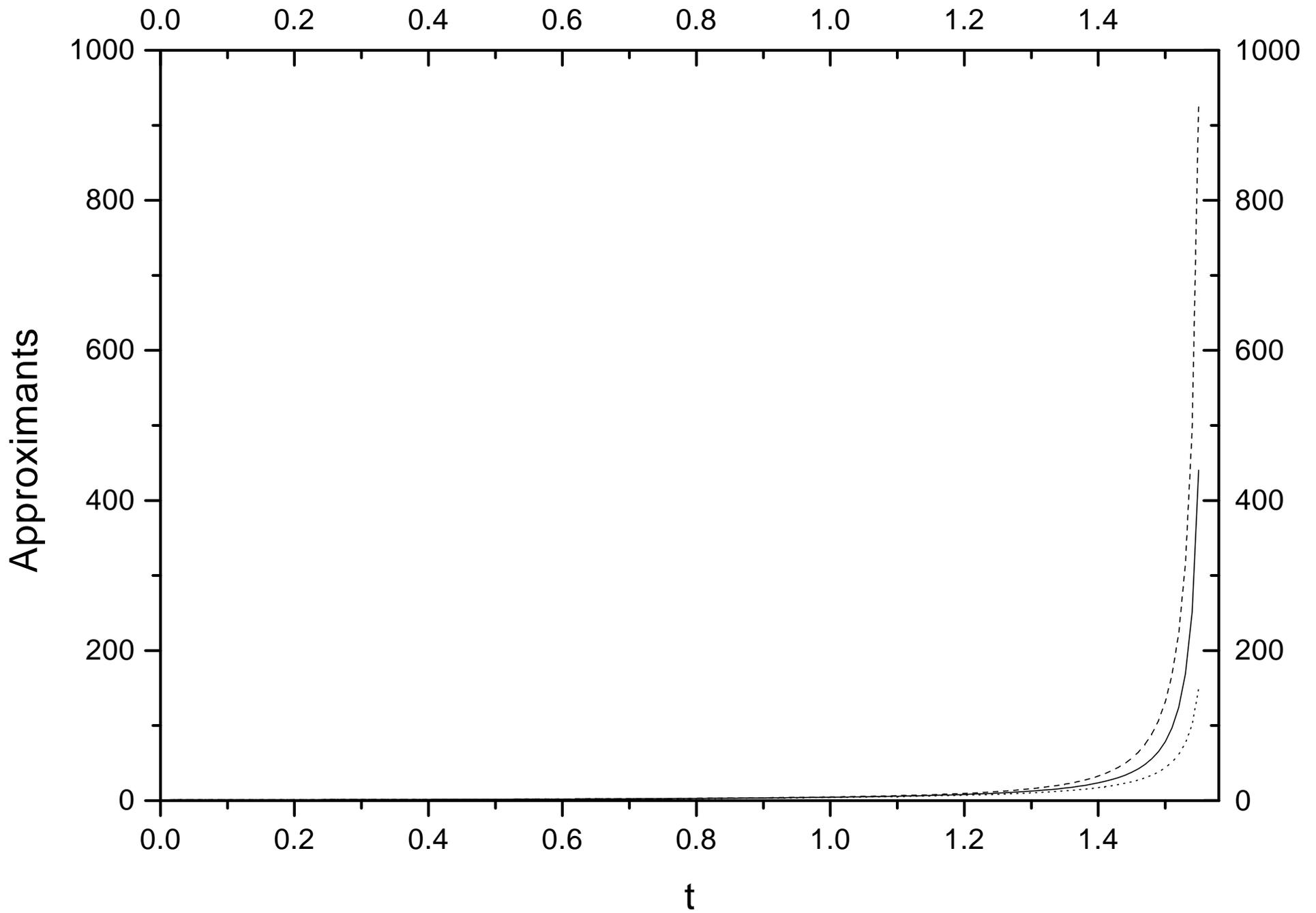
Fig.~7: Same as figure 2 in the regime ${\frac{{F_0}-1}{{F_0}}}~a_1^2<a_2$, $%
F_I>F_0.$

\end{document}